\documentclass[superscriptaddress,nofootinbib,preprintnumbers,showpacs,tightenlines,notitlepage]{revtex4}

\usepackage[latin1]{inputenc}
\usepackage{graphicx}
\usepackage{amssymb}
\usepackage{color}
\usepackage{float}
\usepackage{amsmath}
\usepackage{amsfonts}
\usepackage{dcolumn}
\usepackage{hyperref}
\usepackage{amsthm}
\usepackage{color}
 \newtheorem{prop}{Proposition}

\def\3nab{\tilde{\nabla}}

\def\be {\begin{equation}}
\def\ee {\end{equation}}
\def\ba {\begin{eqnarray}}
\def\ea {\end{eqnarray}}

\newcommand{\bra}[1]{\left(#1\right)}
\newcommand{\bras}[1]{\left[#1\right]}
\newcommand{\brac}[1]{\left\{#1\right\}}

\newcommand{\lc}{\varepsilon}
\newcommand{\lb}{\{}
\newcommand{\rb}{\}}
\newcommand{\E}{{\mathcal E}}
\renewcommand{\H}{{\mathcal H}}
\newcommand{\barray}{\begin{array}}
\newcommand{\earray}{\end{array}}

\newcommand{\hatn}{a}
\newcommand{\dotn}{\alpha}

\newcommand \veps {\varepsilon}

\newcommand{\sss}[1][0.035cm]{\hspace*{#1}}
\newcommand{\A}{{\cal A}}

\newcommand{\dum}{Tweedledum }
\newcommand{\dee}{Tweedledee }
\newcommand{\udot}{{\mathcal A}}
\newcommand{\bea}{\begin{eqnarray}}
\newcommand{\eea}{\end{eqnarray}}
\newcommand{\f}[2]{\textstyle\frac{#1}{#2}}

\newcommand{\udota}{{\cal A}}
\newcommand{\nn}{{\nonumber}}
\renewcommand{\S}{_{\mathsf{S}}}

\begin{document}

 \title{Tidal forces are gravitational waves}
\author{Rituparno Goswami}
 \email{goswami@ukzn.ac.za}
  \affiliation{Astrophysics Research Centre and School of Mathematics, Statistics and Computer Science, University of KwaZulu-Natal, Private Bag X54001, Durban 4000, South Africa.}
 \author{George F. R. Ellis}
 \email{george.ellis@uct.ac.za}
 \affiliation{Department of Mathematics and Applied Mathematics and ACGC, University of Cape Town,
Cape Town, Western Cape, South Africa.}

\begin{abstract}
In this paper we show in a covariant and gauge invariant way that in general relativity, tidal forces are actually a hidden form of gravitational waves. This must be so because gravitational effects cannot occur faster than the speed of light.  Any two body gravitating system, where the bodies are orbiting around each other, may generate negligible gravitational waves, but it is via these waves that non-negligible tidal forces (causing shape distortions) act on these bodies. Although the tidal forces are caused by the electric part of the Weyl tensor, we transparently show that some small time varying magnetic part of the Weyl tensor with non zero curl must be present in the system that mediates the tidal forces via gravitational wave type effects. The outcome is a new test of whether gravitational effects propagate at the speed of light.
\end{abstract}

\pacs{04.20.Cv	, 04.20.Dw}

\maketitle
  
\section{Introduction: Tidal forces in Newtonian gravity}
Let us begin by posing the following question:
\begin{quote}
	{\em How long does it take tidal forces due to the gravitational field of
the Moon to reach the Earth? That is, is there any finite angular lag between the lunar tides in the ocean and the apparent position of the moon in the sky?}
\end{quote}
In Newtonian theory, a tidal force is instantaneously transmitted from the Moon to the Earth due to Newton's Law of Gravitation, but the apparent position of the Moon in the sky lags due to the time it takes light to reach the Earth from the Moon. Hence in that case, there should be an angular lag of 0.66 seconds of arc in the sky between the position of the Moon and the tidal quadrupole on Earth. But a more accurate theory of gravity is General Relativity;  Newtonian theory  is just a first approximation to that  better theory. So the question is, will there be such an angular lag when this is taken into account? The tidal force is mediated by the Weyl tensor, which obeys a wave equation with speed of propagation the speed of light. 
We will show that 
 just like gravitational radiation, tidal forces also propagate at the speed of light, both being special cases of gravitational waves. 
 This allows a new test of whether the gravitational force in fact propagates at the speed of light or not: there should be no such apparent lag. 

\textbf{Tidal forces:} We know that the tidal force is a force that stretches a body towards and away from the centre of mass of another body due to a gradient 
 in the gravitational field of the other body. It is responsible for  tides, tidal locking, tidal disruption of bodies, and the formation of ring systems around celestial bodies.  
Energy transfer between bodies via 
tidal forces is called 
 gravitational induction.
 
  In Newtonian gravity, this phenomenon is governed by the 3-dimensional trace free symmetric tensor defined as follows \cite{Ell71}
\be
E_{ab}=\partial_a\partial_b\Phi-\frac13h_{ab}\partial^c\partial_c\Phi
\label{NewtonialWeyl}
\ee
where $\Phi$ is the gravitational potential, $h_{ab}$ is the metric on 3 dimensional space and the indices $a,b,c$ runs from $1\cdots 3$. This tensor is the Newtonian analogue of the Electric part of the 4 dimensional  Weyl tensor in General Relativity. There is however no Newtonian analogue of the magnetic part of the Weyl tensor \cite{Ell71}, which explains the absence of gravitational waves in Newtonian gravity. 

However the true theory of gravity is relativistic. If this is General Relativity, then \textit{no influence can travel  faster than the speed of light}: the tidal influence cannot be instantaneous in any rest frame. Therefore this influence can either travel along null or timelike curves. To travel along a timelike curve, this influence must be mediated by massive fields. However, in vacuum spacetime no such field is present and in General Relativity, there is no intrinsically defined preferred speed at which such effects would propagate: the only such speed in general relativity \textit{is} the speed of light \cite{c}.  There must be some way in which one can regard tidal forces propagating between astronomical objects in orbital motion, such as the Earth and Moon, or binary pulsars, or even binary black holes, as propagating between them \textit{at} the speed of light, precisely because it is \textit{the} physical speed available. That is, even though we do not usually represent things in this way, in a sense tidal forces are a form of gravitational wave, because otherwise they would not travel at the speed of light. The purpose of this paper is to make  clear how this happens.\\ 

There are several well known examples of energy transfer via gravitational induction in Newtonian gravity:
\begin{enumerate}
\item {\bf The Earth Moon system} \cite{Moon}: The Earth is continuously loosing it's rotational energy through the process of tidal braking caused by the Moon's gravitational field. As a consequence, the length of the day is getting gradually longer by about 2.3 milliseconds per century at the present time. 
This process of Lunar recession results in a net forward acceleration of the Moon along it's orbit and moves the Moon into a slightly larger orbit. There is a steady increase in the average Earth-Moon distance by about 3.8 cm per year.
\item {\bf \dum and \dee thought experiment}: This thought experiment (Bondi \cite{bondi}, McCrea \cite{jvn}) involves two intelligent physically identical spherically shaped creatures (\dum and \dee) who are made up of pliable material that allows them to change their shape. They are instructed to move around each other under their mutual gravitational force along highly eccentric orbits in such a way that when \dum changes shape and \dee responds, their centre of gravity remains fixed in space. Through a series of shape changing operations as dictated by the rules of the game one sees that after each complete rotation, \dum is gaining 
 internal energy as the external tidal force is doing work on him, while \dee is losing 
   internal energy as she is doing work against the external force. This is an excellent example of how the internal energy of a system can be  transferred to another system via {gravitational induction}, in Newtonian gravity.
\end{enumerate}

However, in these situations, Newtonian theory assumes the effect of gravitational induction from one body to another is instantaneous, and hence this violates the principles of special and consequently general relativity. In the subsequent sections, we will recast a two body system in a perturbative formalism within general relativity to transparently examine the mechanism that mediates tidal forces and  energy transfer from one body to another. 

\section{General Relativity, Tidal Forces, Gravitational Radiation and Waves}
We are concerned with tidal forces associated with gravitationally governed orbital motion of two massive bodies, where we know gravitational radiation occurs when seen from outside. At this point we would like to distinguish between the 
terms \textit{Gravitational waves} and \textit{Gravitational radiation} as we will use them.


\textbf{Gravitational waves:} We define the gravitational field as having a wave nature if covariantly defined gravitational variables, such as the electric and magnetic parts of the Weyl tensor or the Regge-Wheeler tensor, obey a {\em closed form wave equation} with the speed of light as the propagation velocity. We apply this definition whether in the context of the near field or the far field.  This is not the same as saying any spacetime with non-zero electric Weyl tensor (which is responsible for tidal deformations) implies the presence of a gravitational wave; that is a necessary but not sufficient condition.  

This definition of a generic gravitational waves is a local concept, defined on any open set $\mathcal{U}$, where the Weyl tensor components or Regge Wheeler tensor solves a closed form wave equation with propagation velocity being equal to the velocity of light in $\mathcal{U}$.
What we transparently show below is that, even if they are first order quantities, the curl of both electric and magnetic Weyl must be non-zero when gravitational waves occur.

\textbf{Gravitational radiation:} Gravitational waves can transfer energy to infinity; this transfer of energy to large distances is \textit{gravitational radiation}, as occurs for example in the case of the Hulse-Taylor binary pulsar \cite{HT}. 
Thus for us, obeying a wave equation is the essence of being a wave; in the far zone (the wave zone) this is  gravitational radiation.  As described in the seminal paper by Newman and Penrose \cite{NP} describing gravitational radiation using 
spin coefficients, for a general asymptotically flat radiative vacuum spacetime, the asymptotic behaviour 
 of the complex Weyl scalar $\Psi_4$ is  $\sim 1/r$. This Weyl scalar describes the gravitational radiation component of the Weyl tensor, and is dominant in the far away radiation zone, where the Riemann tensor is essentially null and all the other Weyl scalars {\em peel off} (or become negligible). The standard definition of gravitational radiation (since the pioneering work of Pirani \cite{Pirani} \cite{Pirani1}) 
is such a propagating curvature tensor, considered in the 
 radiation zone well away from its source.

Thus in our view gravitational radiation is a special case of a gravitational wave, as radiative solutions are indeed  solution of the gravitational wave equations, but for nearly flat background in the vicinity of future null infinity of a future asymptotically simple spacetimes \cite{NP}.  

\textbf{Tidal forces:} These
occur in the near zone in a binary gravitating system, accounting for a time varying tidal deformation in both bodies. But this does not happen instantaneously. 
Rather 
 the gravitational field causing such continuously time varying tidal deformations  obeys a closed form wave equation, and so this is also a case of gravitational waves, as defined above.  Furthermore, this necessarily  implies the presence of non-zero {\em curls of the electric and magnetic Weyl tensor components}, however small they may be. For example, in the Earth-Moon system, due to very slow rotation, the presence of a magnetic Weyl tensor component is negligible in comparison to the electric Weyl tensor component, but it is crucial for the existence of tidal forces that it is non-zero.

\textbf{Other works:}
In the past, there have been numerous works that used retarded relativistic interactions between two bodies to study the near field effects of gravitational interactions, for example \cite{Rad1_Westpfahl} \cite{Rad2_Bel} \cite{Rad3_Damour}  \cite{Rad5}. In these works, it was shown that the near zone solutions have a different character than the wave zone solutions. As described excellently in \cite{damour}, for many years, the Post Newtonian approximation to the first order ($1PN$) was quite an accurate description of {\em many gravitating body problems}, such as the solar system, and various binary stars. However, to take care of strongly gravitating binary systems (like binary black holes and pulsars) one needed to go beyond the ($1PN$) approximations (such as ($2.5PN$)  \cite{Rad4_Schaefer})  to include terms ${\cal O}(v^5/c^5)$ in the equations of motion. To explain the gravitational wave signals emitted binary neutron star  and binary black hole systems, one needed to go even beyond (up to ($4PN$)) \cite{4PN1,4PN2}. In all these studies, the near zone and far zone analysis are quite different. For example, in the approach of Schaefer \cite{Rad4_Schaefer}, we can explicitly see that the tidal tensor (Electric Weyl tensor) decomposes into a Newtonian-like component and a wave like component. In the near zone the first one dominates while the second one dominates in the radiation zone. 

Our view is summarised in \textbf{Figure 1}.

\begin{figure}[h]
	\centering
	\includegraphics[width=0.7\linewidth]{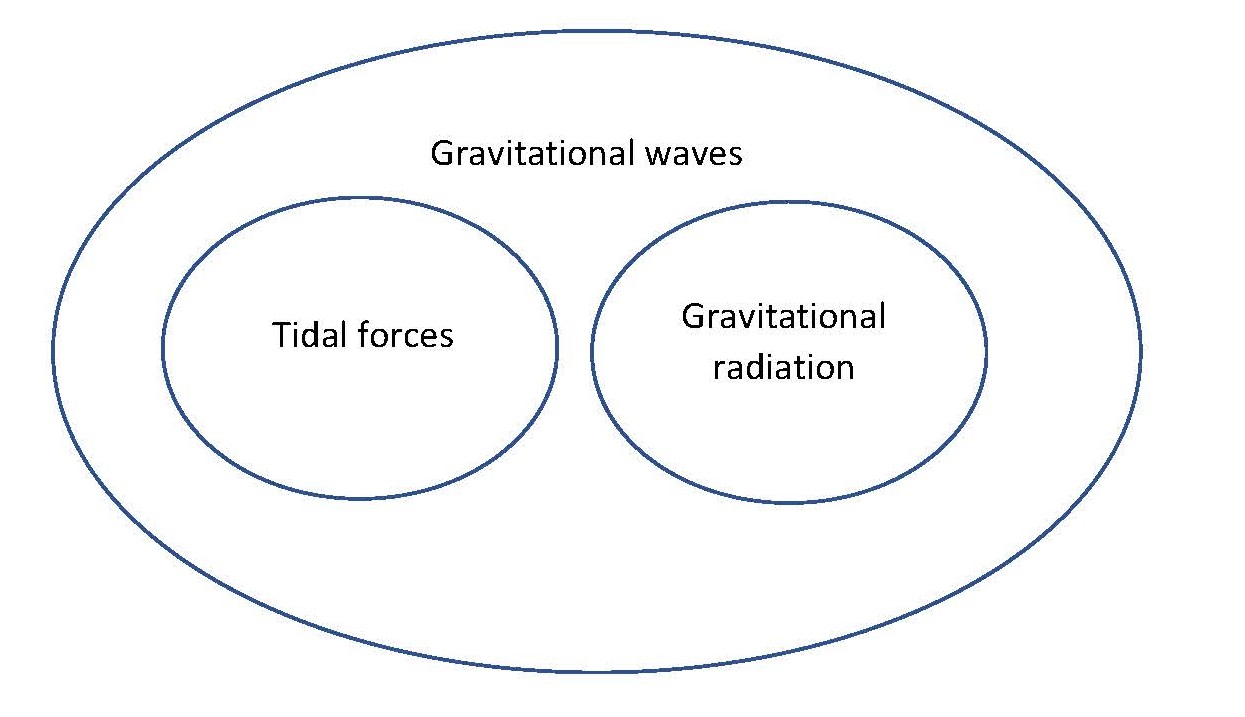}
	\caption[Tidal Forces and Gravitational Radiation are both forms of Gravitational waves]{\textit{Tidal Forces and Gravitational Radiation are both forms of Gravitational waves}}
\end{figure}
\noindent The key points where our work differs from and extends the previous works are as follows:
\begin{enumerate}
\item We present a covariant and gauge invariant closed wave equation that is applicable both in the near and radiation zones. We do not model the fluid bodies themselves: our equations are for vacuum solutions.
 
\item The solutions to these equations indeed have very different characteristics in near and far zones. In the near zone, while the electric Weyl is a zeroth order quantity, in the far zone it is a first order perturbation  quantity. But in both cases, the speed of the gravitational interaction is $c$. 

\item The most crucial part that emerges from our analysis is: whenever there is a time varying tidal interaction, {\em curls of both electric and magnetic Weyl} must be simultaneously non-zero, which is necessary for obtaining a closed wave equation. This is interesting because both the tidal interaction and electric Weyl have a Newtonian counterpart but magnetic Weyl does not. Thus for the propagation of a Newtonian observable, we necessarily need a non-Newtonian variable. 

\item As the near zone tidal effects are a solution to a wave equation with the velocity of propagation equal to that of light, one can easily see that the information of tidal deformations must travel with the velocity of light  between the two bodies. 

\item We are able to derive this result using only a first order perturbative scheme, in contrast to the much more complex perturbative schemes that have been used to derive much more comprehensive results. 

\item This leads to a potential test as to whether the speed of tidal interaction is the speed of light. This amounts to a test as to whether general relativity is the correct theory of gravitation.
\end{enumerate}
We acknowledge that some of this may be implicit in the above mentioned literature. We believe it is useful to have it made explicit in this simple direct way.

\section{The Weyl Tensor as the free gravitational field }

It is common to think of the metric tensor  $g_{ab}(x^j)$ as the gravitational field in general relativity, with  
the Einstein Field Equations 
\begin{equation}\label{eq:EFE}
G_{ab}:= R_{ab}-\frac{1}{2}R g_{ab}= \kappa T_{ab} + \Lambda g_{ab}
\end{equation}
being second order partial differential equations for $g_{ab}$ where $R_{ab}$ is the Ricci tensor determined by $g_{ab}$, $R$ the Ricci scalar, $T_{ab}$ the matter stress-energy tensor, and $\Lambda$ the cosmological constant. In the case of the local propagation of gravitational waves through a vacuum, the cosmological constant is irrelevant and $T_{ab}=0$ so (\ref{eq:EFE}) reduces to
\begin{equation}\label{eq:EFE_vac}
 R_{ab}-\frac{1}{2}R g_{ab}= 0 \Leftrightarrow R_{ab} = 0,   
\end{equation}
which in the weak field limit is well known to have wave solutions \cite{LL}.\\

However an alternative viewpoint \cite{Ell71} \cite{Covariant} is to view the Weyl tensor
\begin{equation}\label{eq:Weyl}
C_{abcd} :=R_{abcd} + \bra{g_{a[d}R_{c]b}+g_{b[c}R_{d]a}}+\f13Rg_{a[c}g_{d]b}\;, 
\end{equation}
as the free gravitational field, and the metric tensor as its (2nd order) potential field. This tensor has the same symmetries as the Riemann tensor but in addition is traceless: 
\begin{equation}
C^a_{\,\,\,bad}= 0;
\end{equation}
in the case of a vacuum, it is identical to the Riemann tensor:
\begin{equation}\label{eq:Weyl_vacuum}
\{R_{ab} = 0\} \Leftrightarrow \{C_{abcd} =R_{abcd}\}
\end{equation}

Now the useful thing is this: given a timelike vector field $u^a(x^j): u_a u^a = -1$ representing a family of fundamental observers, in exact parallel to the way one can split the electromagnetic field $F_{ab}$ into electric and magnetic parts $E_a$, $H_b$  in the rest frame of $u^a$ ($E_a u^a = 0$, $H_a u^a = 0$) one can similarly split $C_{abcd}$ into electric and magnetic parts $E_{ab}$, $H_{ab}$ (\ref{E},\ref{H}); but this time they are symmetric traceless tensors orthogonal to $u^a$ (see appendix (\ref{A1} for details):  
\begin{eqnarray}\label{key}
E_{ab} = E_{(ab)}, \,E^a_{\,\,\,a}= 0, \, \, E_{ab}u^a =0,\\ H_{ab} = H_{(ab)}, \,H^a_{\,\,\,a}= 0,\,\,H_{ab}u^a =0
\end{eqnarray}
It is the electric part $E_{ab}$ of the Weyl tensor that is responsible for tidal forces, because it enters the geodesic deviation equation and so causes relative displacement of freely falling particles, and more generally it causes relative displacement of particles in elastic media, which is how tidal forces are generated and gravitational waves are detected \cite{Pirani}. (for effect of Newtonian gravity on elastic media see \cite{mora}).

Now just as the Maxwell Equations \begin{equation}\label{eq:maxwell}
F_{[ab;c]} = 0,\,\,F^{ab}_{\,\,\,\,\,\,;b}=0
\end{equation}  give the usual $\dot{E}$, $curl E$, $\dot{H}$, $curl H$  equations in the source free case (where ($\dot{}$) denotes the directional derivative along the timelike congruence), in the vacuum case (\ref{eq:EFE_vac}), the Bianchi identities
\begin{equation}\label{eq:Bianchi}
R_{ab[cd;e]} = 0 \Leftrightarrow R^a_{\,\,bcd;a}= 0
\end{equation}
(the equivalence holding only in 4 dimensions) lead to the temporal and spatial derivatives of electric and magnetic part of Weyl tensor as follows, in the vacuum case (\ref{eq:EFE_vac}): 
\begin{eqnarray} 
\dot{E}^{<ab>}-\epsilon^{cd<a}D_cH^{b>}_d=-\Theta E^{ab}+3\sigma^{<a}_cE^{b>}_c\nn\\
+\epsilon^{cd<a}\bra{2A_cH^{b>}_d+\omega_cE^{b>}_d}\;,\label{dotE}
\end{eqnarray}
\begin{eqnarray} 
\dot{H}^{<ab>}+\epsilon^{cd<a}D_cE^{b>}_d=-\Theta H^{ab}+3\sigma^{<a}_cH^{b>}_c\nn\\
-\epsilon^{cd<a}\bra{2A_cE^{b>}_d-\omega_cH^{b>}_d}\;,\label{dotH}
\end{eqnarray}
\begin{eqnarray} 
D_bE^{ab}-3\omega_bH^{ab}-\epsilon^{abc}\sigma_{bd}H^d_c=0\;,\label{divE}
\end{eqnarray}
\begin{eqnarray} 
D_bH^{ab}+3\omega_bE^{ab}+\epsilon^{abc}\sigma_{bd}E^d_c=0\;. \label{divH}
\end{eqnarray}
Here $D$ is the projected covariant derivative operator on 3-space orthogonal to the timelike congruence and angle brackets denote the projected trace free part.\\

The key point is the following: gravitational waves arise by taking the dot derivative of the dot-E equation, commuting the dot and curl operators on H, and substituting from the H-dot equation to obtain a wave equation for $E_{ab}$, where the wave speed is the speed of light c \cite{Hawking}.
In empty space with a non-expanding congruence $u^a$, this reduces to the usual form of
the linearized theory \cite{Hawking}:
\be\label{eq:wave}
\Box E_{ab} = 0 .
\ee
Similarly we get a wave equation for $H$, precisely as in the EM case. The key result is
 \begin{quote}
 	\textbf{Gravitational wave criterion} 	
 	\textit{We only get a wave equation for $E_{ab}$
 		if}
 		 \begin{equation}\label{eq:criterion}
 		 curl H \neq 0 \Rightarrow H_{ab} \neq 0.
 		 \end{equation} 
	\end{quote}
 Hence in the kind of tidal system represented by the Earth-Moon system, even though we normally think of it in Newtonian like terms with tidal forces represented by the  $div E$ equations, and  $H_{ab}$ either zero or certainly negligible, \textit{$H_{ab}$ cannot be neglected when there are tidal forces}. However unlike the usual gravitational wave case where $E_{ab}$ and $H_{ab}$ are oscillating out of phase at high frequency and with equal amplitudes, in this case  \textit{$E_{ab}$ and $H_{ab}$ are oscillating out of phase at low frequency and $E_{ab}$ has a much greater amplitude than $H_{ab}$, which has a non-zero curl}. This is what enables the tidal force field to propagate from one body to another at the speed of light.\\
 
There is one further thing to note. At any instant in the comoving reference frame, the $div E$ equation is a Laplacian equation representing an instantaneous influence whereby $E_{ab}$ can be regarded as traveling instantaneously from the Earth to the Moon. However this is not in fact instantaneous propagation, as this equation is a constraint that was required to be satisfied in order to set up consistent initial data way back in the past (when the relevant structures were formed). It remains true today because conservation of the constraint equations is a consistency condition for the whole set of propagation equations to be true. That is, they remain true at a later time because they were true at an earlier time and were propagated forward in time via the time development equations. Their specific form at any epoch is because wave propagation took place and kept them true at later times, 
having initially been true.\\
 
The visualisation of the interaction of $E$ and $H$ during the emission of gravitational radiation, due to the wavelike nature of their interaction, is beautifully illustrated by Thorne and co-workers \cite{Thorne1}, \cite{Thorne2}. However they do not relate this to tidal forces as we do here. The fully covariant 1 + 3 electromagnetic analogy for gravity is also developed in \cite{Maartens}.  We remark that given that $H$ may become large during late stages of inspiral of neutron stars, it is possible that tidal forces due to $E$ will be  augmented by gravitoelectromagnetic effects relating $H$ to rotation of fluid masses  (see for example \cite{wiki}). We will not pursue that issue here.

\section{Two body system in General Relativity: A perturbative approach}

Unlike Newtonian gravity, the Einstein equations do not have a general solution for a two body system. Therefore we resort to a perturbative approach, where the presence of the second body is taken as a perturbation over the spacetime geometry generated by the first one. We perform our entire analysis using a semitetrad covariant $1+1+2$ formalism, generated by a timelike and a preferred spacelike congruence (please see appendix \ref{A1} and \ref{A2} for a complete description of the formalism and important geometrical definitions and identities). This covariant formalism helps in recasting the perturbed field equations in a gauge invariant way that brings out the gauge invariant results.
\subsection{The background spacetime}
For simplicity and transparency in our calculations, we assume that the central body is spherically symmetric. Then the vacuum spacetime around this central body must be Schwarzschild, by Birkhoff's theorem \cite{birk,AlmBir}. In that case the geometry is necessarily static, and the directional derivatives of all geometrical variables along the timelike congruence must vanish. Thus the only non-zero geometrical variables (\ref{Dgeom}) in the background spacetime are \cite{chris1}:
\be\label{D0}
\mathcal{D}_0=\{ \phi,\udot, \E\},
\ee
that satisfies the following propagation equations and constraints
\ba
\hat\phi&=&-\frac12\phi^2 -\E\;, \label{phihat}\\
\hat\E&= &-\frac32\phi\E\;,\label{Ehat}\\
\hat\A&= &-\bra{\A+\phi}\A\;,\label{Ahat}\\
\E&=&-\A\phi\;.
\ea
Furthermore, we can write the Gaussian curvature of the 2-sheets perpendicular to timelike and preferred spacelike congruences (see appendix \ref{A2} for details) as 
\be 
K = - \E + \frac14 \phi^{2} \;.
\label{GaussCurv} 
\ee
From the above equations it is clear that the electric part of the Weyl scalar is proportional to a $(3/2)th$ power of the Gaussian curvature and the proportionality constant (that is the Schwarzschild mass $m$) sets up a covariant scale in the problem. We can also define the areal radius of the two sheets $r$, such that the Gaussian curvature is $(1/r^2)$. In that case we must have
\be\label{r}
r=\left(- \E + \frac14 \phi^{2}\right)^{-\frac12}\;.
\ee
The propagation equations can now be integrated in terms of this variable and we get \cite{chris1}
\ba
\phi=\frac{2}{r}\sqrt{1-\frac{2m}{r}}\;&,&\;\A=\frac{m}{r^2}
\left[1-\frac{2m}{r}\right]^{-\frac12}\\
\E=-\frac{2m}{r^3}\;&,&\;K=\frac{1}{r^2} 
\label{Schw1} 
\ea
\subsection{The perturbed spacetime}
Let us now consider a second spherical body, whose Schwarzschild mass is much smaller than the mass of the central body (which is the covariant scale in the problem), which is rotating around the central body. The situation is very similar to the Earth-Moon system, where the ratio of moon's mass to that of Earth's mass is  $\approx 0.1$. The spacetime around and in between these bodies will then be perturbed Schwarzschild, and all the geometrical quantities that vanished in the background will now be non-zero but their magnitude will be much smaller than the invariant scale. 

The set of  first order quantities are 
 \cite{clarksonlrs}
\ba
\mathcal{D}_1&=&\left\{ \Theta, \Sigma, \Omega, \H, \xi,
\A^{a},\Omega^{a}, \Sigma^{a}, \alpha^a, \right. \nonumber\\ 
&&\left. a^a, \E^{a}, \H^{a}, \Sigma_{ab}, \E_{ab}, \H_{ab},\zeta_{ab}
\right\} \,. 
\ea
We can now write the first order equations for the system as follows.
The time evolution equations for 
$\xi$ and $\zeta_{\lb ab\rb}$ are 
\be 
\dot\xi =\bra{\udota-\f12\phi}\Omega
+\f12\lc_{ab}\delta^a\dotn^b+\f12\H\;, \label{dotxinl} 
\ee 
\bea
\dot\zeta_{\lb ab\rb}&=&\bra{\udota-\f12\phi}\Sigma_{ab} +\delta_{\lb
a}\alpha_{b\rb} -\lc_{c\lb a}\H_{b\rb}^{~~c}\;. \label{dotzetanl}
\eea 
The vorticity evolution equations are  
\bea
\dot\Omega&=&\f12\lc_{ab}\delta^a\udota^b+\udota\xi \;,
\eea 
\bea
\dot\Omega_{a}+\f12\lc_{ab}\hat\udota^b&=&\f12\lc_{ab}\bra{-\udota
    a^b+\delta^b\udota-\f12\phi\udota^b}\;.
\eea
The shear evolution equations are 
\bea\label{replace1}
\dot\Sigma-\f23\hat\A=\f13(2\A-\phi)\A-\f13\delta_a\A^a-\E
\eea
\bea
\dot\Sigma_{\lb ab\rb}&=&\delta_{\lb a}\udota_{b\rb}
+\udota\zeta_{ab} -\E_{ab}\;,
\eea
\bea
\dot\Sigma_{a}-\f12\hat\udota_{a}&=&\f12\delta_a\udota+\bra{\udota-\f14\phi}\udota_a
    \nonumber\\&&
    +\f12\udota a_a-\f32\Sigma\alpha_a-\E_a\;.
\eea
Evolution equation for $\hat e_a$ are:
\bea
\hat\alpha_{a}-\dot a_{a}&=&\bra{\f12\phi-\udota}\bra{\Sigma_a+\lc_{ab}\Omega^b}\nonumber\\
&&-\bra{\f12\phi+\udota}\alpha_a-\lc_{ab}\H^b.\label{hatalphanl}
\eea
Electric Weyl evolution is:
\bea\label{replace2}
\dot\E=\bra{\f32\Sigma-\Theta}\E+\lc_{ab}\delta^a\H^c
\eea
\bea
\dot\E_{a}+\f12\lc_{ab}\hat\H^b&=&
    \f34\lc_{ab}\delta^b\H+\f12\lc_{bc}\delta^b\H^c_{~a}
    -\f34\E(\Sigma_a+2\alpha_a)\nonumber\\&&
    +\f34\E\lc_{ab}\Omega^b
     -\bra{\f14\phi+\udota}\lc_{ab}\H^b\;,
\eea
\bea
\dot\E_{\lb ab\rb}-\lc_{c\lb a}\hat\H_{b\rb}^{~~c}&=&
    -\lc_{c\lb a}\delta^c\H_{b\rb}
    -\f32\E\Sigma_{ab} \nonumber\\&&
    +\bra{\f12\phi+2\udota}\lc_{c\lb a}\H_{b\rb}^{~~c}\;.
\eea
Magnetic Weyl evolution is:
\bea
\dot\H&=&-\lc_{ab}\delta^a\E^b-3\xi\E\;,
\eea
\bea
\dot\H_{a}-\f12\lc_{ab}\hat\E^b&=&-\f32\E\lc_{ab}\udota^b 
    +\f34\E\lc_{ab}a^b-\f12\lc_{bc}\delta^b\E^c_{~a}\nonumber\\&&
    +\bra{\f14\phi+\udota}\lc_{ab}\E^b
   -\f34\lc_{ab}\delta^b\E\;,
\eea

\bea
\dot\H_{\lb ab\rb}+\lc_{c\lb a}\hat\E_{b\rb}^{~~c}&=&
    +\f32\E\lc_{c\lb a}\zeta_{b\rb}^{~~c}
    -\bra{\f12\phi+2\udota}\lc_{c\lb a}\E_{b\rb}^{~~c}\nonumber\\&&
     +\lc_{c\lb a}\delta^c\E_{b\rb}\;.
\eea
Sheet expansion evolution is given by:
\bea\label{replace3}
\dot\phi&=&\bra{\f23\Theta-\Sigma}\bra{\A-\f12\phi}+\delta_a\alpha^a
\eea
The Raychaudhuri equation is
\bea\label{replace4}
\hat\A-\dot\Theta=-\delta_a\A^a-\bra{\A+\phi}\A
\eea
The
propagation equations of $\xi$ and $\zeta_{\lb ab\rb}$ are: 
\bea
\hat\xi&=&-\phi\xi +\f12\lc_{ab}\delta^aa^b\;,
\eea 
\bea \hat\zeta_{\lb ab\rb}&=&-\phi\zeta_{ab} +\delta_{\lb
a}\hatn_{b\rb } -{\cal
E}_{ab}\;.
\label{hatzetanl} 
\eea 
The shear divergence is given by : 
\bea\label{sigmahat}
\hat\Sigma-\f23\hat\Theta&=&-\f32\phi\Sigma-\delta_a\Sigma^a
\eea
\bea
\hat\Sigma_{a}-\lc_{ab}\hat\Omega^b&=&\f12\delta_a\Sigma
    +\f23\delta_a\theta - \lc_{ab}\delta^b\Omega-\f32\phi\Sigma_a
     -\f32\Sigma a_a\nonumber\\&&
    +\bra{\f12\phi+2\udota}\lc_{ab}\Omega^b
    -\delta^b\Sigma_{ab}\;,
\eea
\bea
\hat\Sigma_{\lb ab\rb}&=&\delta_{\lb a}\Sigma_{b\rb} -\lc_{c\lb a}\delta^c\Omega_{b\rb}
    -\f12\phi\Sigma_{ab}\nonumber\\&&
    +\f32\Sigma\zeta_{ab}-\lc_{c\lb a}\H_{b\rb}^{~~c}\;.
\eea
The vorticity divergence equation is:
\bea
\hat\Omega&=&-\delta_a\Omega^a+\bra{\udota-\phi}\Omega\;.\label{hatOmSnl}
\eea
The Electric Weyl divergence is
\bea\label{replace5}
\hat\E&=&-\delta_a\E^a-\f32\phi\E\;,
\eea
\bea
\hat\E_{a}&=& \f12\delta_a\E -\delta^b\E_{ab}
      -\f32\E a_a-\f32\phi\E_a\;.
\eea
The Magnetic Weyl divergence is:
\bea
\hat\H&=& -\delta_a\H^a
    -\f32\phi\H-3\E\Omega\;,
\eea
\bea
\hat\H_{a}&=& \f12\delta_a\H-\delta^b\H_{ab}
    -\f32\E\lc_{ab}\Sigma^b
    +\f32\E\Omega_a\nonumber\\&&
    +\f32\Sigma\lc_{ab}\E^b
    -\f32\phi\H_a\;.
\eea
The sheet expansion propagation is:
\bea\label{replace6}
\hat\phi&=&-\f12\phi^2+\delta_aa^a-\E\;.
\eea
We also have the following constraints:
\bea
\delta_a\Omega^a+\lc_{ab}\delta^a\Sigma^b=\bra{2\A-\phi}\Omega+\H~,
\eea
\bea
\frac12\delta_a\phi-\veps_{ab}\delta^b\xi-\delta^b\zeta_{ab}&=&-\E_a~,
\label{divzetanl1}
\eea
\bea
\delta_a\Sigma-\frac23\delta_a\Theta+2\veps_{ab}\delta^b\Omega +2\delta^b\Sigma_{ab}&=&
    -\phi\bra{\Sigma_a-\veps_{ab}\Omega^b}\nn\\&& -2\veps_{ab}\H^b . 
    \label{divSigmanl1}
\eea

\section{Making the system gauge invariant}
The equations in the previous section have both zeroth order and first order terms. The rules for mapping the zeroth order terms from background spacetime manifold to the perturbed spacetime manifold defines the gauge choice. However quantities that vanish in the background spacetime are automatically gauge invariant by the Stewart and Walker lemma \cite{SW,EllisBruni}. Therefore to make the system of equations gauge invariant we define three new variables that vanish in the background \cite{chris,anne}:
\be
\mathcal{D}_{GI}=\left\{W_a =\delta_a \E,  Y_a =\delta_a \phi, Z_a =\delta_a \A \right\}
\ee
The evolution and propagation equations for the new variables are    
\bea
\dot W_a &=&\frac32 \phi \,\E \bra{\alpha_a + \Sigma_a - \veps_{ab} \Omega^b}
+ \frac32 \E\bra{ \delta_a \Sigma - \frac23 \delta_a \Theta}  \nn \\
&&+\, \veps_{bc} \delta_a \delta^b \H^c 
\label{newEdot} ~, 
\eea
\bea
\dot Y_a &=& \bra{\frac12 \phi^2 + \E}\bra{\alpha_a + \Sigma_a - \veps_{ab} \Omega^b}
+ \delta_a\delta_c \alpha^c \nn \\
&&+\,\bra{\frac12 \phi - \A}\bra{\delta_a \Sigma - \frac23 \delta_a \Theta}   ~, 
 \label{newphidot} 
\eea
\bea 
 \hat W_a &=& -\,2 \phi \,W_a - \frac32 \E\, Y_a 
+ \frac32 \phi \,\E \,\hatn_a - \delta_a \delta_b \E^b~, 
\label{newEhat} 
\eea
\bea
\hat Y_a &=& -\, W_a - \frac32 \phi\, Y_a +\bra{\frac12 \phi^2 + \E} a_a  
+ \delta_a \delta_b a^b, 
\label{newphihat} 
\eea
\bea
\hat Z_a &=& -\bra{\frac32 \phi + 2 \A} Z_a - \A\, Y_a + \A \bra{\phi + \A} a_a
 \nn\\
&&+\delta_a \dot \Theta - \,\delta_a \delta_b \A^b  ~.  
\label{newAhat}
\eea
These equations add no new information to what has already been given in 
the previous section, however they are now gauge invariant. We can now replace (\ref{replace1}) by (\ref{newAhat}), (\ref{replace2}) by (\ref{newEdot}), (\ref{replace3}) by (\ref{newphidot}), (\ref{replace5}) by (\ref{newEhat}) and (\ref{replace6}) by (\ref{newphihat}). This will make the complete system of equations gauge invariant. The following additional constraints are obeyed by the new variables:
\bea
\veps_{ab} \delta^aW^b &=& 3 \phi\, \E \,\xi ~,
\label{Wconst}
\\
\veps_{ab} \delta^aY^b&=& \bra{\phi^2 + 2 \E} \xi ~,
\label{Yconst}
\\
\veps_{ab} \delta^aZ^b &=& 2\A \bra{\phi + \A} \xi ~.
\label{Zconst}
\eea
It is also useful to replace (\ref{replace4}) with
 \bea
\delta_a\dot\Sigma-\frac23\delta_a\dot\Theta&=&-\,W_a-\A \,Y_a
-\phi\, Z_a-\delta_a\delta_b\A^b ~.
\label{WYZ}
\eea
Introducing these new variables eliminates the study of possible spherically symmetric perturbations 
(for  they 
not represented by  these variables). However since by Birkhoff's theorem, all the vacuum  
spherically symmetric static spacetimes are Schwarzschild, we do not lose any true degrees 
of freedom by omitting them. 

\section{Regge Wheeler tensor and wave equation}

As shown extensively in \cite{chris, anne}, from the first order traceless tensors on the two sheets, we can construct a dimensionless, covariant, gauge invariant, transverse tracefree tensor $M_{\{ab\}}$ in the following way:
\be
M_{ab}=\frac12\phi \,r^2\,\zeta_{ab}-\frac13r^2\,\E^{-1}\,\delta_{\lb a}W_{b\rb}\;.
\label{Mab}
\ee
Provided $\veps_{bc}\delta_a\delta^b {\cal H}^c \neq 0$, in parallel to the way that (\ref{eq:wave}) follows provided $curl H \neq 0$, this tensor obeys the wave equation
\be
\ddot M_{\lb ab\rb}-\hat{\hat M}_{\lb ab\rb}-\A\,{\hat M}_{\lb ab\rb}
+\bra{\phi^2 + \E}M_{ab}-\delta^2 M_{ab} =0~.
\label{RMtensorwave}
\ee
The tensor $M_{ab}$ is known as {\em Regge Wheeler tensor} and the wave equation (\ref{RMtensorwave}) dictates both the odd and even parity perturbations. It is interesting to note that the tensor $M_{ab}$ gives a measure of sheet deformation, via the electric part of Weyl scalar and the deformation tensor related to the preferred spacelike direction. At this point we would like to emphasize two key points:
\begin{enumerate}
\item {\bf All information regarding tidal forces are encoded in gravitational waves}. We know that the tidal forces between the two bodies are a consequence of the electric part of the Weyl tensor ($\E$ in this case). Therefore the evolution and propagation equations of the variable $\E$ should give the complete description of these forces. However, we have already seen that the evolution and propagation equations of $W_a\equiv \delta_a \E$, carries the same information. That is the reason why we could replace \ref{replace2}) by (\ref{newEdot}) and (\ref{replace5}) by (\ref{newEhat}). 

Therefore the tensor $M_{ab}$ carries all the information about  tidal forces. Unlike Newtonian gravity, where the effects of tidal forces are instantaneous, in GR these effects travel via the gravitational waves described by (\ref{RMtensorwave}).
\item {\bf Manifestation of tidal forces requires non-zero curl of magnetic Weyl}: One of the necessary conditions for the existence of the wave equation (\ref{RMtensorwave}), is that the magnetic part of Weyl tensor along with it's curl must be strictly non vanishing in the perturbed spacetime (although these quantities may be small {\em i.e.} of the first order). Therefore for the transmission of the action of a zeroth order non negligible electric part of Weyl, we must need, at least to the first order, the presence of the magnetic part. The existence of this non-zero curl of $H$, and hence of non-zero $H$, can be explicitly seen from the equation (\ref{newEdot}). When we take a time derivative of that equation and use the commutation relations of time and sheet derivatives for the term involving $curl H$, we get a closed form wave equation (\ref{RMtensorwave}). If curl  $H$ vanishes in the perturbed spacetime, we will not get a wave equation for the Regge Wheeler tensor. 
\end{enumerate}

\section{Nature of waves that mediates tidal forces}

In the previous section we have already established that, in a two body system, the action of tidal forces travels  from one body to another via the gravitational wave equation (\ref{RMtensorwave}). In other words, these tidal deformations are gravitational waves. To investigate in more depth the nature of the gravitational waves mediating tidal effects, we decompose the geometrical variables as an infinite sum of components relative to a basis of harmonic functions. 
This will enable us to replace angular derivatives appearing in the equations by a harmonic coefficient. We follow \cite{chris1} where the harmonics were introduced in a covariant manner. 
Introduce the set of dimensionless spherical harmonic functions $Q=Q^{(\ell,m)}$, 
with $m=-\ell,\cdots,\ell$, defined in the background, as eigenfunctions 
of the spherical Laplacian operator such that 
\be
\delta^2 Q =-\, \frac{\ell(\ell+1)}{r^2} \,Q~,
\label{SH}
\ee
where $Q$ is covariantly constant, $\hat{Q} = 0 = \dot{Q}$. 
As we are interested in the transmission of the effects of the scalar $\E$, we expand the wave equation in terms of the scalar harmonics in the following way:
We expand any first order
scalar ${\Psi}$ in terms of the harmonic functions as
\be
{\Psi}=\sum_{\ell=0}^{\infty}\sum_{m=-\ell}^{m=\ell} {\Psi}\S^{(\ell,m)}
Q^{(\ell,m)} = {\Psi}\S \,Q,
\ee
where the sum over $\ell$ and $m$ is implicit in the last equality. We use the
subscript $\mathsf{S}$ to remind us that ${\Psi}$ is a scalar, and that a
spherical harmonic expansion has been made. Due to the spherical symmetry 
of the background, we can drop $m$ in the equations.\\
\\
The replacements  which must be made for scalars when expanding the equations 
in spherical harmonics are
\bea
 {\Psi} &=& {\Psi}\S\, Q ~,                
  \\
 \delta_a {\Psi} &=& r^{-1}{\Psi}\S \,Q_a~,
 \\
 \veps_{ab}\delta^b {\Psi} &=& r^{-1}{\Psi}\S\, \bar Q_a ~,
\eea
where the sums over $\ell$ and $m$ are implicit.
We can expand \ref{RMtensorwave} in scalar harmonics as
\bea
\ddot { M}-\hat{\hat{  M}}-\A\,\hat { M}
+\bras{\frac{\ell\bra{\ell+1}}{r^{2}}+3\E}{ M}=0~,\label{RW}
\eea
In appropriate coordinates the wave equation \ref{RW} is the usual {\it Regge\,-Wheeler
equation} that appears in many GR textbooks. As physically expected, the effects of tidal forces must be determined by the small values of the multipole moment `$\ell$'. For example, in the Earth-Moon system, $\ell=1$ will account for the maximum effect of  tidal forces. For more massive neutron star binaries, however, this value can go up to $\ell=4$or $5$ to cover the complete tidal distortions as specified by the Tidal Love Number (due to the other body) and the Rotational Love Number (due
to the rotation of each body) \cite{Tanja}. \\

Apart from this, for any nearly spherical vacuum (which definitely fits for the vacuum region between the Earth and the Moon, for example), we have the {\em Almost Birkhoff theorem} \cite{AlmBir}, which states that for an almost spherically symmetric vacuum spacetime there
always exists a vector in the local $[u,e]$ plane which almost
solves the Killing equations. If this vector is timelike then the
spacetime is locally almost static, and if the Killing vector is
spacelike the spacetime is locally almost spatially homogeneous. \\

As a direct consequence of this, when the time derivative
all the background quantities are zero, we can easily see that the time derivatives of
the first order quantities at a given point is of the same order of smallness as
themselves. Hence the first order quantities still remain  ``small'' as the time evolves. Therefore the time variation of the gravitational waves will be slow as set by the time scale of the rotation of the binary system around each other. This definitively
 indicates the following:\\

\begin{prop}
The tidal forces experienced by the bodies in a binary system are very slow time varying (given by the time scale of rotation) gravitational waves of small multipole moments, with a large electric part of Weyl and a small magnetic part of Weyl, but both with non vanishing curl: thus
\begin{equation}\label{eq:curlcurl}
		| curl E| |curl H| \neq 0.
\end{equation}

\end{prop}


\section{
	Tidal Force proof that $\textbf{v}_{grav}=\textbf{c}$}

So the answer to the question at the beginning of the paper is that it takes tidal forces around 1.3 seconds to reach the Earth from the Moon, because they travel at the speed of light. Can we detect this? Yes we can. The point is that this is precisely the same as the time it takes the light that forms our image of the Moon to arrive. Hence we should see the centre of the Moon to be precisely aligned with the direction of the tidal force, both having taken the same time to reach the Earth. This is a null result: any deviation represents a difference in propagation speeds of light and the gravitational force.    
\begin{quote}\textbf{Tidal Force Test of $v_{grav}=c$ as predicted by General Relativity} 
\textit{The direction of tidal attraction on Earth should be perfectly aligned to the observed position of the Moon in the sky (neglecting the much smaller solar tide effect). 
}\end{quote}
This will then be a definite proof that the effect of tidal forces travels at the speed of light. As the light from the moon takes 1.3 seconds to reach Earth, if the effect of the tides was instantaneous as predicted in Newtonian gravity, there would be an angular lag of 0.66 seconds of arc between the position of the Moon and the tidal force (since the position of the Moon we see at a given instant is where it was 1.3 seconds before). Therefore no lag will definitely imply that the information of tidal forces travels with the speed of light. 
In contrast, if the true theory of gravity involved a massive graviton traveling at less than the speed of light, there would be a lead in the effect instead of a lag.

\subsection{Detecting the tidal force}
One could either try this via the Earth itself, or via a test body in a laboratory, or by laser ranging of near Earth satellites.

\paragraph{Earth's tidal deformation} To measure the Earth's tidal deformation with required accuracy is  non-trivial and requires highly sensitive laser extensometers, water tube tiltmeters, ocean bottom and land spring gravimeters and SG (superconducting gravimeters) \cite{Tide1,Tide2} . Also satellite based Synthetic aperture radar (SAR) interferometry experiments are in progress to accurately measure the Earth's Crust Deformation \cite{SAR}. However there is a further problem: one would then have to do a complex simulation of the effect of the Moon's tidal force on the Earth with its complex interior structure, oceans of varying depths, and continental structures. This may well mean that this method is impractical in practice.
Using the Earth's tidal deformation involves a complex inhomogeneous body that is far from ideal. 

\paragraph{Laboratory Detection} A second option might be a laboratory detection of the tidal force. This could for example use a metallic sphere or shell of say 2 meter diameter with either passive or active detection of the tidal force quadrupole, and hence its major axis. Passive detection would use either strain gauges, as proposed by Weber long ago, or laser interferometry, perhaps in the interior of a shell. Active detection would use activators on the surface of the sphere that would keep it perfectly spherical as measured by laser interferometry, the magnitude of the force needed measuring the quadrupole and hence detecting its direction. 

We know that the tidal acceleration is given by:
\be
{\vec {a}}_{t,{\text{axial}}}\approx \pm {\hat {r}}~2\Delta r~G~{\frac {M}{R^{3}}}
\ee

\noindent Now if we consider a ball of radius $0.5 m$ then $\Delta r = 1 m$ and the tidal acceleration due to earth's gravity will be of the order of $6 \times 10^{-6} m/s^2$ (i.e in the order of a micrometer per second${}^2$)
and the tidal acceleration due to the Moon will be of the order of $1.8 \times 10^{-13}m/s^2$ which is even less than a picometer per second${}^2$. Regrettably this does not look feasible.  

\paragraph{Laser Ranging of near Earth Satellites} This seems to be the most viable option. Satellites can be tracked with great accuracy, and indeed that is happening all the time, specifically with the GPS network. Amongst the forces perturbing these orbits will be the Moon's gravitational field, but also of course anisotropies in the Earth's gravitational field and tidal forces due to the Sun. However the periodicity of the Moon's tidal force should make it easy to identify. We suspect that this might give the required accuracy and enable the test to be carried out in practice.

\subsection{Detecting the position of the Moon}
 
To detect the alignment of the position vector of the moon (with respect to the centre of the earth) and direction of tidal deformation of the earth, one can use the data from Lunar Laser Ranging (LLR) experiment, that can measure the distance to the moon with millimetre precision \cite{LLR}.

\subsection{Verifying gravitational force propagation speed}
 If one could achieve the required accuracy of  measurement that can measure an angle of the order of an arcsecond between these vectors, we would be able to experimentally confirm that indeed the position of the moon and earths lunar tidal deformations are exactly aligned. This would then  confirm that  tidal effects travel with the speed of light, riding on a near field gravitational wave, as predicted by General Relativity. This would then imply that gravitational radiation should also travel at the speed of light.
 
 We concede that the practicality of doing this to the required precision is problematic. Nevertheless it may  be  useful to set it as an interesting challenge to our observational colleagues.


\section{Conclusion}
In hindsight, it is not surprising that a non-zero magnetic Weyl tensor is associated with tidal forces, for we know that any binary system emits gravitational waves, detectable at ``infinity'' (i.e. at large distance from the system \cite{Finite_inf} \cite{Finite_inf1}), that carry away mass and energy from the system \cite{HT}. We would like to emphasize here again, that in this investigation we only deal with tidal forces that are associated with gravitationally governed orbital motion of two massive bodies, where we know gravitational radiation occurs when seen from outside. Hence even though a static spherical body \textit{per se } has no magnetic Weyl tensor component (its Weyl tensor is type D), it generates non-zero $H$ and $curl H$ when in orbit about another such body.\\

Thus it should not be surprising that the same is true \textit{within} the system, and the tensor $E_{ab}$, traveling at the speed of light because of the wave equation (\ref{eq:wave}),  exerts influences between the two bodies which they experience as tidal forces. Our argument then is that even though rapid speeds of bodies and large distances are not involved, this should be regarded as a form of low frequency gravitational waves, in which the electric part of the Weyl tensor dominates the magnetic part; but it is crucial for the story as a whole that the latter is non-zero. Our criterion for existence of a gravitational wave (via a closed form wave equation) is (\ref{eq:curlcurl}), that is  
		\be\label{eq:criterion1}
		| curl E| |curl H| \neq 0
		\ee
		because that is the criterion for existence of a wave equation (\ref{eq:wave}) for $E$, and so determines that it is in fact propagating at the speed of light (if (\ref{eq:criterion}) and  (\ref{eq:criterion1}) are not true, such an equation does not follow from the $\dot E$, $\dot H$ equations.) \\
		
The next obvious question would be, how small is small? or what should be the relative magnitude of the magnetic and the electric Weyl? If we approximate the Earth Moon system by a Kerr spacetime (we emphasize here, that this a only an approximation as the system is definitely not Kerr), then up to the first order in the rotation parameter $a$ of Kerr metric, we have
\be
\frac{\H}{\E}\approx \frac{a}{r}
\ee
The equation above gives a measure of the {\it smallness} of the magnetic Weyl with respect to the electric Weyl.\\
 
  \begin{acknowledgments}
  	We thank John Miller and  Rainer Weiss for helpful discussions.\\
  	
RG thanks the South African National Research Foundation (NRF) for support, and GE thanks the University of Cape Town Research Committee (URC) for financial  support.
 \end{acknowledgments}

\appendix

\section{Semitetrad 1+3 formalism}\label{A1}

In the 1+3 formalism \cite{Covariant}, the timelike unit vector ${u^{a}}$ ${\left(u^{a}\sss u_{a} = -1\right)}$ is used to split the spacetime locally in the form ${\mathcal{R} \otimes \mathcal{V}}$, where ${\mathcal{R}}$ is the timeline along ${u^{a}}$ and ${\mathcal{V}}$ is the 3-space perpendicular to ${u^{a}}$. Thus the metric becomes
\begin{equation}
 g_{ab} = - u_{a}\sss u_{b}+ h_{ab},
\end{equation}
where ${h_{ab}}$ is the metric on 3-space perpendicular to $u^a$. The covariant time derivative along the observers' worldlines, denoted by `${\sss\sss^{\cdot}\sss\sss}$', is defined using the vector ${u^{a}}$, as
\begin{equation} 
\dot{Z}^{a ... b}{}_{c ... d} = u^{e}\sss\nabla_{e}\sss Z^{a ... b}{}_{c ... d},
\end{equation} 
for any tensor ${Z^{a...b}{}_{c...d}}$. The fully orthogonally projected covariant spatial derivative, denoted by `\sss ${D}$\sss', is defined using the spatial projection tensor ${h_{ab}}$, as
\begin{equation}
D_{e}\sss Z^{a...b}{}_{c...d} = h^r{}_{e}\sss h^p{}_{c}\sss...\sss h^q{}_{d}\sss h^a{}_{f}\sss...\sss h^b{}_{g}\sss\nabla_{r}\sss Z^{f...g}{}_{p...q},
\end{equation}
with total projection on all the free indices. The covariant derivative of the 4-velocity vector ${u^{a}}$ is decomposed irreducibly as follows
\begin{eqnarray}
\nabla_{a}\sss u_{b} &=& -u_{a}\sss A_{b} + \frac{1}{3}h_{ab}\sss\Theta + \sigma_{ab} + \varepsilon_{abc}\sss \omega^{c},
\end{eqnarray}
where ${A_{b}}$ is the acceleration, ${\Theta}$ is the expansion of ${u_{a}}$, ${\sigma_{ab}}$ is the shear tensor, ${\omega^{a}}$ is the vorticity vector representing rotation and ${\varepsilon_{abc}}$ is effective volume element in the rest space of the comoving observer. Furthermore the energy momentum tensor of matter, decomposed relative to ${u^{a}}$, is given by
\begin{eqnarray} \label{3.Tab}
T_{ab} &=& \mu\sss u_{a}\sss u_{b} + p\sss h_{ab} + q_{a}\sss u_{b} + u_{a}\sss q_{b} + \pi_{ab},
\end{eqnarray}
where ${\mu}$ is the effective energy density, ${p}$ is the isotropic pressure, ${q_{a}}$ is the 3-vector defining the heat flux and ${\pi_{ab}}$ is the anisotropic stress.
We write down the definitions of important components including the kinematical, Weyl and matter quantities in the 1+3 formalism.
Angle brackets denote orthogonal projections of vectors onto the three space as well as the projected, symmetric and trace-free part of tensors.
\begin{eqnarray}
v_{<a>}& = &h^{b}{}_{a}\sss\dot{V}_{b},\\ 
Z_{<ab>}& =& \bra{h^{c}{}_{(a}\sss h^{d}{}_{b)} - \frac{1}{3}h_{ab}\sss h^{cd}}\sss Z_{cd}.
\end{eqnarray}

\begin{eqnarray}
\varepsilon_{abc} &=& \sqrt{|\textrm{det } g|}\sss\delta^0{}_{[a}\sss \delta^1{}_{b}\sss\delta^2{}_{c}\sss\delta^3{}_{d]} \sss u^{d}, \\
\varepsilon_{abc}\sss\varepsilon^{def} &=& 3! \sss h^d{}_{\left[ a \right.} \sss h^e{}_{b} \sss h^f{}_{\left. c \right]}, \\
\varepsilon_{abc}\sss\varepsilon^{dec} &=& 2\sss h^d{}_{\left[ a \right.} h^e{}_{\left. b\right]}, \\
A_{b} &=& \dot{u_{b}}, \\ 
\Theta &=& D_{a}\sss u^{a}, \\ 
\sigma_{ab} &=& \left(h^{c}{}_{(a}\sss h^{d}{}_{b)} - \frac{1}{3}h_{ab}\sss h^{cd}\right)D_{c}\sss u_{d}, \\
\omega^{a} &=& \varepsilon^{abc}\sss D_{b}\sss u_{c}, \\
E_{ab} &=& C_{abcd}\sss u^{c}\sss u^{d} = E_{<ab>}, \label{E}\\
H_{ab} &=& \frac{1}{2} \varepsilon_{ade}\sss C^{de}{}_{bc}\sss u^{c} = H_{<ab>}, \label{H}\\
\mu &=& T_{ab}\sss u^{a}\sss u^{b}, \\ 
p &=& \frac{1}{3}\sss h_{ab}\sss T^{ab}, \\
q_{a} &=& q_{<a>} = - h^{c}{}_{a}\sss T_{cd}\sss u^{d},\\
\pi_{ab}&=& \left(h^{c}{}_{(a}\sss h^{d}{}_{b)} - \frac{1}{3}h_{ab}\sss h^{cd}\right)T_{cd}.
\end{eqnarray}

\section{Semitetrad 1+1+2 formalism}\label{A2}

In the 1+1+2 formalism \cite{chris1}, the 3-space ${\mathcal{V}}$ is now further split by introducing the unit vector ${e^{a}}$ orthogonal to ${u^{a}}$ ${\left(e^{a}\sss e_{a} = 1, u^{a}\sss e_{a} = 0\right)}$. The 1+1+2 covariantly decomposed spacetime is given by 
\begin{equation}\label{2.Nab}
g_{ab} = -u_{a}\sss u_{b} + e_{a}\sss e_{b} + N_{ab},
\end{equation}
where ${N_{ab}}$ ${\left(e^{a}\sss N_{ab} = 0 = u^{a}\sss N_{ab}, N^{a}{}_{a} = 2\right)}$ projects vectors onto 2-spaces called `sheets' , orthogonal to ${u^{a}}$ and ${e^{a}}$.  We introduce two new derivatives for any tensor ${\phi_{a...b}{}^{c...d}}$:
\begin{eqnarray}
\label{hatderiv}
\hat{\phi}_{a...b}{}^{c...d} &\equiv& e^{f}\sss D_{f}\sss \phi_{a...b}{}^{c...d}, \\
\label{deltaderiv}
\delta_{f}\phi_{a...b}{}^{c...d} &\equiv& N_{f}{}^{j} N_{a}{}^{l} ... N_{b}{}^{g} N_{h}{}^{c} ... N_{i}{}^{d}  D_{j}\phi_{l...g}{}^{h...i}.
\end{eqnarray}
 
The  1+3 kinematical quantities and anisotropic fluid variables are split irreducibly as
\begin{eqnarray}
A^{a} &=& \mathcal{A}\sss e^{a} + \mathcal{A}^{a}, \\
\omega^{a} &=& \Omega\sss e^{a} + \Omega^{a}, \\
\sigma_{ab} &=& \Sigma\left(e_{a}\sss e_{b} - \frac{1}{2}\sss N_{ab}\right) + 2\sss\Sigma_{(a}\sss e_{b)} + \Sigma_{ab}, \\
\label{2.qa} q_{a} &=& Q\sss e_{a} + Q_{a}, \\
\label{2.pia} \pi_{ab} &=& \Pi\left(e_{a}\sss e_{b} - \frac{1}{2}\sss N_{ab}\right) + 2\sss\Pi_{(a}\sss e_{b)} + \Pi_{ab}.
\end{eqnarray}
 
The fully projected 3-derivative of ${e^{a}}$ is given by
\begin{eqnarray}
D_{a}\sss e_{b} &=& e_{a}\sss a_{b} + \frac{1}{2}\sss\phi\sss N_{ab} + \xi\sss\varepsilon_{ab} + \zeta_{ab},
\end{eqnarray}
where traveling along ${e^{a}}$, ${a_{a}}$ is the sheet acceleration, ${\phi}$ is the sheet expansion, ${\xi}$ is the vorticity of ${e^{a}}$ (the twisting of the sheet) and ${\zeta_{ab}}$ is the shear of ${e^{a}}$ (the distortion of the sheet). 

Any 3-vector ${\Phi^{a}}$ can be irreducibly split into ${\chi}$, a scalar component along ${e^{a}}$, and a 2-vector ${\chi^{a}}$, which is a sheet component orthogonal to ${e^{a}}$, as follows
\begin{eqnarray} \label{decomp1}
\Phi^{a} = \chi\sss e^{a} + \chi^{a}  
\end{eqnarray}
where $\chi \equiv \Phi_{a}\sss e^{a}$ and  $ \chi^{a} \equiv N^{ab}\sss \Phi_{b} \equiv \Phi^{\bar{a}}$.
Similarly we can split a projected, symmetric, trace-free tensor ${\Phi_{ab}}$ into scalar, 2-vector and 2-tensor parts as follows
\begin{equation} \label{decomp2}
\Phi_{ab} = \Phi_{<ab>} = \chi\left(e_{a}\sss e_{b} - \frac{1}{2}\sss N_{ab}\right) + 2\sss\chi_{(a}\sss e_{b)} + \chi_{ab},
\end{equation}
where 
\begin{eqnarray}
\chi &\equiv& e^{a}\sss e^{b}\sss \Phi_{ab} = -N^{ab}\sss\Phi_{ab}, \nonumber\\
\chi_{a} &\equiv& N_{a}{}^{b}\sss e^{c}\sss\Phi_{bc}, \nonumber\\
\chi_{ab} &\equiv& \chi_{\brac{ab}} = \left(N_{(a}{}^{c}N_{b)}{}^{d} - \frac{1}{2}N_{ab} N^{cd}\right)\Phi_{cd}.
\end{eqnarray}
The curly brackets denote the part of the tensor that is projected, symmetric and trace-free on the sheet.
We write down the definitions of important components in the 1+1+2 formalism.
\begin{eqnarray}
\varepsilon_{ab} &\equiv& \varepsilon_{abc}\sss e^{c} = \sqrt{|\textrm{det } g|}\delta^0{}_{[a}\delta^1{}_{b}\delta^2{}_{c}\delta^3{}_{d]}e^{c} u^{d}, \\
E_{ab} &=& \mathcal{E} \left(e_{a}\sss e_{b} - \frac{1}{2}\sss N_{ab}\right) + 2\sss\mathcal{E}_{(a}\sss e_{b)} + \mathcal{E}_{ab}, \\
H_{ab} &=& \mathcal{H}  \left(e_{a}\sss e_{b} - \frac{1}{2}\sss N_{ab}\right) + 2\sss\mathcal{H}_{(a}\sss e_{b)} + \mathcal{H}_{ab}
\end{eqnarray} 
\begin{eqnarray}
\sigma^{2} &=& \frac{1}{2}\sigma_{ab}\sigma^{ab} = \frac{3}{4}\Sigma^{2} + \Sigma_{a}\Sigma^{a} + \frac{1}{2}\Sigma_{ab}\Sigma^{ab}, \\
a_{a} &\equiv& e^{c}\sss D_{c}\sss e_{a} = \hat{e}_{a}, \\
\alpha_a&\equiv&N^c_a\dot{e_c},\\
\phi &\equiv& \delta_{a}\sss e^{a}, \\
\xi &\equiv& \frac{1}{2}\sss\varepsilon^{ab}\sss\delta_{a}\sss e_{b}, \\ 
\zeta_{ab} &\equiv& \delta_{\{a}\sss e_{b\}}.
\end{eqnarray}

The 1+1+2 split of the full covariant derivatives of ${u^{a}}$ and ${e^{a}}$ are as follows
\begin{eqnarray} \label{2.delAUb}
\nabla_{a}\sss u_{b} &=& -u_{a}\left(\mathcal{A}\sss e_{b} + \mathcal{A}_{b}\right) + e_{a}\sss e_{b} \left(\frac{1}{3}\sss\Theta + \Sigma \right)\nonumber\\
 &&+ e_{a}\left(\Sigma_{b} + \varepsilon_{bc}\sss\Omega^{c}\right) + \left(\Sigma_{a} - \varepsilon_{ac}\sss\Omega^{c}\right) e_{b}\nonumber\\
 &&+ N_{ab}\left(\frac{1}{3}\sss\Theta - \frac{1}{2}\sss\Sigma\right) + \Omega\sss\varepsilon_{ab} + \Sigma_{ab}, \\
\nabla_{a}\sss e_{b} &=& -\mathcal{A}\sss u_{a}\sss u_{b} - u_{a}\sss\alpha_{b} + \left(\frac{1}{3}\sss\Theta + \Sigma \right)e_{a}\sss u_{b} \nonumber\\
&&+ \left(\Sigma_{a} - \varepsilon_{ac}\sss\Omega^{c}\right)u_{b}  + e_{a}\sss a_{b}\nonumber\\
 &&+  \frac{1}{2}\sss\phi\sss N_{ab}+ \xi\sss\varepsilon_{ab} + \zeta_{ab}.		
\end{eqnarray}

We can now immediately see that the Ricci identities and the doubly contracted Bianchi identities, that specifies the evolution of the complete system, can now be written as the time evolution and spatial propagation and spatial constraints of a irreducible set of geometrical and thermodynamic variables.

The irreducible set of geometric variables
\begin{eqnarray}
\label{Dgeom}
\mathcal{D}_{geom}
 = 
\{\Theta, \sss \mathcal{A}, \sss\Omega, \sss\Sigma, \sss\mathcal{E}, \sss\mathcal{H}, \sss\phi, \sss\xi, \sss\mathcal{A}_{a}, \sss\Omega_{a}, \sss\Sigma_{a},\\ 
\sss\alpha_{a}, \sss a_{a}, \sss\mathcal{E}_{a}, 
\sss\mathcal{H}_{a}, \sss\Sigma_{ab}, \sss\zeta_{ab}, 
\sss\mathcal{E}_{ab}, \sss\mathcal{H}_{ab}\}
\nonumber
\end{eqnarray}
together with the irreducible set of thermodynamic variables
\begin{eqnarray}\label{Dtherm}
\mathcal{D}_{therm}&=&\brac{\mu, \sss p, \sss Q, \sss \Pi, \sss Q_{a}, \sss \Pi_{a}, \sss \Pi_{ab}},
\end{eqnarray}
make up the key variables in the 1+1+2 formalism.

\end{document}